\documentclass[journal]{IEEEtran}
\makeatletter\if@twocolumn\PassOptionsToPackage{switch}{lineno}\else\fi\makeatother

\usepackage{graphicx}
\usepackage[T1]{fontenc}
\ifCLASSINFOpdf
\else
\fi
\usepackage{amsmath}
\usepackage[cmintegrals]{newtxmath}
\usepackage{url,multirow,morefloats,floatflt,cancel,tfrupee}
\makeatletter

\AtBeginDocument{\@ifpackageloaded{textcomp}{}{\usepackage{textcomp}}}
\makeatother
\usepackage{colortbl}
\usepackage{xcolor}
\usepackage{pifont}
\usepackage[nointegrals]{wasysym}
\makeatletter

\def\mcWidth#1{\csname TY@F#1\endcsname+\tabcolsep}

\def\cAlignHack{\rightskip\@flushglue\leftskip\@flushglue\parindent\z@\parfillskip\z@skip}
\def\rAlignHack{\rightskip\z@skip\leftskip\@flushglue \parindent\z@\parfillskip\z@skip}

\@ifundefined{etal}{}{}

\usepackage{ifxetex}
\ifxetex\else\if@twocolumn\@ifpackageloaded{stfloats}{}{\usepackage{dblfloatfix}}\fi\fi

\AtBeginDocument{
\expandafter\ifx\csname eqalign\endcsname\relax
\def\eqalign#1{\null\vcenter{\def\\{\cr}\openup\jot\m@th
  \ialign{\strut$\displaystyle{##}$\hfil&$\displaystyle{{}##}$\hfil
      \crcr#1\crcr}}\,}
\fi
}

\AtBeginDocument{%
  \@ifpackageloaded{endfloat}%
   {\renewcommand\efloat@iwrite[1]{\immediate\expandafter\protected@write\csname efloat@post#1\endcsname{}}}{\newif\ifefloat@tables}%
}%

\def\BreakURLText#1{\@tfor\brk@tempa:=#1\do{\brk@tempa\hskip0pt}}
\let\lt=<
\let\gt=>
\def\processVert{\ifmmode|\else\textbar\fi}

\@ifundefined{subparagraph}{
\def\subparagraph{\@startsection{paragraph}{5}{2\parindent}{0ex plus 0.1ex minus 0.1ex}%
{0ex}{\normalfont\small\itshape}}%
}{}

\newcommand\role[1]{\unskip}
\newcommand\aucollab[1]{\unskip}
  
\@ifundefined{tsGraphicsScaleX}{\gdef\tsGraphicsScaleX{1}}{}
\@ifundefined{tsGraphicsScaleY}{\gdef\tsGraphicsScaleY{.9}}{}
\def\checkGraphicsWidth{\ifdim\Gin@nat@width>\linewidth
	\tsGraphicsScaleX\linewidth\else\Gin@nat@width\fi}

\def\checkGraphicsHeight{\ifdim\Gin@nat@height>.9\textheight
	\tsGraphicsScaleY\textheight\else\Gin@nat@height\fi}

\def\fixFloatSize#1{}
\let\ts@includegraphics\includegraphics

\def\inlinegraphic[#1]#2{{\edef\@tempa{#1}\edef\baseline@shift{\ifx\@tempa\@empty0\else#1\fi}\edef\tempZ{\the\numexpr(\numexpr(\baseline@shift*\f@size/100))}\protect\raisebox{\tempZ pt}{\ts@includegraphics{#2}}}}

\AtBeginDocument{\def\includegraphics{\@ifnextchar[{\ts@includegraphics}{\ts@includegraphics[width=\checkGraphicsWidth,height=\checkGraphicsHeight,keepaspectratio]}}}

\DeclareMathAlphabet{\mathpzc}{OT1}{pzc}{m}{it}

\def\URL#1#2{\@ifundefined{href}{#2}{\href{#1}{#2}}}

\def\UrlOrds{\do\*\do\-\do\~\do\'\do\"\do\-}%
\g@addto@macro{\UrlBreaks}{\UrlOrds}

\edef\fntEncoding{\f@encoding}

\makeatother

\newif\ifmultipleabstract\multipleabstractfalse%
\usepackage{tabulary}
\makeatletter
\AtBeginDocument{\@ifpackageloaded{longtable}{%
\def\LT@makecaption#1#2#3{%
  \LT@mcol\LT@cols c{\hbox to\z@{\hss\parbox[t]\LTcapwidth{%
    \sbox\@tempboxa{#1{#2: } #3}%
    \ifdim\wd\@tempboxa>\hsize
      #1{#2: }\textsc{#3}%
    \else
      \hbox to\hsize{\hfil\box\@tempboxa\hfil}%
    \fi
    \endgraf\vskip\baselineskip}%
  \hss}}}
}{}}
\makeatother

\makeatletter
\let\citep\cite
\let\citet\cite
\makeatother

\begin{document}

%

        \title{All one needs to know about shared micromobility simulation: a complete survey}
      
\author{Yixuan Liu, Yuhan Tang, and Yati Liu}

\maketitle 

\begin{abstract}
As the shared micromobility becomes a part of our daily life and environment, we expect the number of low-speed modes for first-and-last mile trips to grow rapidly. The shared micomobility is expected to serve billions of humans, bringing us considerable advantages. With this growth, shared micromobility simulation such as docked stations based shared bikes, dockless shared bikes and e-scooters, are regarded as promising solutions to deal with a large number of first-and-last mile trips. In this paper, we first provide a comprehensive overview of shared micromobility simulation and its related validation metrics. Next, we classify the research topics of shared micromobility simulation, summarize, and classify the existing works. Finally, challenges and future directions are provided for further research.
\end{abstract}
    
%
\IEEEpeerreviewmaketitle

\section{Introduction}
Micromobility - a revolutionary transportation strategy - is defined as tiny, light and even portable vehicles running at a speed below 15 mph. It is now regarded as a promising way to solve the ``last mile'' problem. Despite the existing solution like walking to transit and driving to transit, micromobility is a less tiring and more economical-friendly alternative. Therefore, micromobility is placed high hopes for reducing the reliance on cars for traveling less than 10 miles. They are easy to ride, require less space, cause almost no pollution and perform well in short-distance house-to-house transportation. If developed well, micromobility will even cause voluntary behavioral change [1] and innovate the whole transportation system.

Although there are already a lot of researches about physical road \citep{huang2019chemical}, driving \citep{chai2021automated} \citep{everts2021making} and aviation \citep{bauranov2021quantifying}, the researches on aviation are still rare. Everything new is the unity of forward and tortuous. Competing for road space with other means of transportation, micromobility makes transportation systems much more complex. We can't simply classify them as pedestrians, bicycles or cars. What changes may happen when micromobility {\textendash} a new mode of transportation {\textendash} is added? Does it affect people's mode choice or route choice? Is it more energy-efficient? The previous way of simulation is no longer suitable for micromobility. What can be called an innovative mesoscopic micromobility simulation? Therefore, in order to have a better understanding on micromobility and give full play to its strengths, a systematic micromobility stimulation is necessary.

Researches on the simulation of micromobility range from microscopic to macroscopic and increasingly attention is paid to human behavior and decision-making. Wu J L C[2] is one of the first researchers to simulate micromobility in 1996 using the Sun Microsystems Mobile IP model.  Thanks to the invention of the Multi-Agent Transport Simulation Toolkit (MATSim), researches on the simulation of micromobility flourish. To test whether micromobility is feasible and safe enough to be a new kind of transportation, researchers concentrate on microscopic simulation targeting one single vehicle. Alberto Fern\'{a}ndez, et al. establish a microscopic agent-based model, where different types of users can be defined and adjust according to private objectives, to test different station capacities, station distributions, and balancing strategies.[3] After the wide spread of micromobility in San Francisco Bay area, the focus sfhift to mesoscopic simulation to find out passengers' choice on a larger scale. Focusing on the impacts of different policies of incentives, Alberto L\textasciiacute opez Santiago, et al. build and validate the model.[4] In order to evaluate and compare different alternatives for the system designs Francesc Soriguera, et al. include rebalancing/reposition dynamics and battery constraints for e-bikes in their simulation.[5] In addition to the models mentioned above, previous research contributions also summarize the potential of micromobility in literature reviews. Abduljabbar R L, et al. consolidates knowledge on how micro-mobility meet sustainability outcomes in urban environments through their review.[6] In summary, the research emphasis of micromobility is turning from traffic flow to people's behavior.

Different from the papers mentioned above, the contribution of this paper is three-fold: (1) We propose an exhaustive taxonomy of concepts in the area of micromobility simulation, and present a comprehensive survey on micromobility simulation. (2) We provide a comprehensive review on micromobility simulation, how it is defined, and how it is related to or different from other similar simulation paradigms, such as transit simulation, ride and hail simulation, car simulation and pedestrian simulation. (3) We have compiled a list of challenges and future directions for research in micromobility simulation.
    
\section{Methodology}

\subsection{Taxonomy of micromobility simulation}\textbf{*Taxonomy for inputs}

Inputs are the files or data that are put in, taken in, and operated on before the micromobility simulation, which determines the basic framework of the simulation model as well as part of operational rules.  It should be noted that inputs are not limited to variables, files in any formats or even independent variables can both be inputs, for the import of which is a prerequisite for model operation. For micro-mobility simulation, basic required inputs are: configuration files, building the data structure and containing optional parameters to formulate the operational rules; network files, describing the features of the transportation network; environmental files, taking into account the geographical factors(i.e. terrain, elevation, climate, etc) and social factors(such as population, occupation, land use, communities, etc); travel demand files, stimulating and determining the features of travel; behavioral rules files, formulating the travel routes, directions and options. 

\textbf{* Taxonomy for parameters}

Parameters decide both the data structure and system structure for micromobility simulation, and formulate the rules to limit the way that the elements in models can be done. All the operation and behaviors have to work within the parameters that had already been set. According to the parameter setting purposes and the roles of parameters played in the simulation, six categories of parameters can be found. However, intersections exist for the usage of parameters can be found in every part of the system.

\begin{itemize}
  \item \relax \textbf{Parameters for configuration }are the parameters what are usually contained in configuration files as one of the elements of inputs, determining the structure both for data and system as well as basic operational rules. 
  \item \relax \textbf{Parameters for service level }represent the efficiency and availability of the model for evaluation the simulation, such as the availability of bikes.
  \item \relax \textbf{Parameters for network and route design} determine the detailed structure of the transportation network and route selection formulations.
  \item \relax \textbf{Parameters for constraint} set constraints for the operation and behaviors, such as the charging and discharging rules and time worth. The same parameter may be set in different ways depending on the researches' considerations.
  \item \relax \textbf{Parameters for utility functions} take many factors such as time worth and cost into consideration in order to evaluate the traveling utility.
\end{itemize}
  The term micro-mobility, in the sense of taking short-trips with small vehicles, is widely believed in related researches to contain three specific vehicles: bikes, e-bikes and e-scooters. The National Association of City Transportation Officials defines shared micromobility as: ``all shared-use fleets of small, full or partially human-powered vehicles such as bikes, e-bikes, and e-scooters.'' Bike sharing system including bikes and e-bikes exists two common modes, station-based mode requires bikes to be rented or returned on the docking stations while bikes can be locked optionally in the service-area in dockless mode\unskip~\cite{1219518:23250173,1219518:23250134}. The e-bike sharing system and e-scooter sharing system consist of e-vehicles, batteries powered by lithium ion, automatic locking rack and a station for rental and charging. Micro-mobility makes short trips easier and low-cost, allowing people healthy, fast, emission-free and effective movement. Also, it will bring environmental and social benefits for alleviating the problems caused by traffic pollution and congestion. However, the randomness, asymmetry and peculiarity of the micro-mobility demands in different areas cause operating agencies many problems, such as the system unbalance(A simulation model for public) or even system failure. 

As a positive solution to these problems and predict the potential usage, micro-mobility simulation emerges, and the research scope of which can be classified into three levels, including the microscopic level, the mesoscopic level and macroscopic level. Compared to the formulation of management policy representing the macroscopic simulation, microscopic simulation owns much utility for simulating the behaviors and interplay-performance of individual commuters, but the size of the research areas are usually limited. Mesoscopic simulation has a much larger research area, which is usually a metropolitan region, focusing on users equilibrium and their usage of utility functions. Therefore, the results of the mesoscopic simulation could be more persuasive and reliable. 

Establishing a system should take many influential things into consideration, like the system itself(number, location and density of the station, number of the vehicles, size of the areas...), demand gereration rules, users generation, rebalacing strategies, and environmental factors such as weather and policy. Due to the additional elements of the e-vehicles sharing system, many operational complications should be introduced into the original sharing system simulation, such as charging rate, battery constraints et al. Shuguang Ji et al.(2014) developed a Monte Varlo simulation model of e-bike sharing system to evaluate the performance\unskip~\cite{1219518:23250172}.

Generally, for micro-mobility simulation in the light of framework there exists two common categories, strategy-based simulation, which focuses on the mechanisms or elements of the system[-], and agent-based simulation, which emphasizes the interactions between involved agents whatever passive or positive, and more consideration towards behavioral rules, influential factors and conformity with the reality is given. All the elements of the system in agent-based simulation is established as an operable unit, which can be adjusted according to the patterns and schemes. The adjustment of one agent will affect variables and even cause action alternation of other agents, that is, the decision of one agent will generate an interaction in the simulation system. MATSIM is a software developed for agent-based transportation simulation, with which some reproducation of the micro-mobility sharing system has been realized. Cornelia Hebenstreit Hebenstreit[4] et al. built a dynamic bike sharing module on MATSIM with the implement of a within day rescheduling and choice probability to provide improvement to the current system\unskip~\cite{1219518:23250171}.

\subsection{Inputs {\textemdash} What were the inputs for the simulation? What were the commonalities and differences across different methods? Follow the above organization.}Before conducting the simulation, models need to be configured such that the elements in the model as well as the behavioral rules meet the requirement. This is what the inputs work for.  As explained in the taxonomy, inputs are the files or data that are put in the model to formulate the whole system, which can be independent variables and files in any formats. The following text provides an explication of the basic input files with regard to the micro-mobility simulation models as well as explanation of the encompassed parameters.  

\textbf{Configuration files }providing the configuration options to construct the data structure as well as model structure.  For MATSim, which is a multi-agent and activity-based simulation framework commonly used for micro-mobility modeling and conducting adaptive and co-evolutionary algorithms, a config.xml file is needed to determine how the simulation behaves with a settings list. It contains all configuration parameters which are divided into groups according to the modules. For example, the group related to network contains the parameter inputNetworkFile; the group related to the Controller includes the parameter firsIteration and lastIteration.  

\textbf{Network file} \textbf{\space }consisting of nodes and links, providing the routes for travelers to move around. A network.xml file is a necessity in MATSim to describe the road transportation network by claiming the features of the nodes and links. For the nodes, it will provide its identity as well as coordinate values of x and y. For the links, the identity, starting nodes (e.g., from), ending nodes (e.g., to), length (typically in meters), capacity (number of vehicles that traverse the link, typically in vehicle per hour), free speed (the maximum allowed speed of vehicles, typically in meters per second), perm lanes (the number of available lanes in the direction from the staring nodes and ending nodes, all the links are uni-directional), and modes of the links (chosen from car, bike, and taxi). 

Other models on transport network need different kind of inputs. The bi-level programming model used by Miller et al., (1992) [14] contains the basic parameters for determining the structure of the network: fearture-related parameters, such as the denoting and number of the arcs, paths, nodes with their sets; flow-related parameters, such as the flow on a specific path or arc; cost-related parameters, such as the unit cost of transportation on a specific path, the cost incurred by the locating firm to ship, and the total variable cost of production; demand-and-supply-related parameters, such as the inverse demand(or supply) function and the quantity supplied at a specific node; and other parameters. 

\textbf{Area files} containing all the distribution of agents, stops, racks, routes can be clearly seen against the area file. For example, MATSim needs an ESRI file in Shapefile format to provide detailed population demographics and land-use patterns. 

\textbf{Population.xml}, containing the necessary information about the model for MATSim. Population is the full set of agents, describing the day plans of agents in a hierarchical structure (as follows) to generate travel demand. The minimum example contains a list of persons, and each person has a plan with a score attribute (obtained in the scoring stage after simulation), and each plan posses a list of activities, legs as well as routes. If the simulation starts, the link will be automatically assigned to the nearest activity and a route for a leg will be figure out. 

The activities and legs providing the agents' actions in each plan: activities has three parameters: the type of the traveling(home, work and others), the link and the end time(except for the last activity). End time is sometimes not required because a duration replaces it in some activities which are automatically generated by routing algorithms[7]. Link will explicate the location from which the activity is reached by providing the link or an x and y coordinate value. Link is required in MATSim so Controller will figure out and use the nearest link to the given point if the link attribute is missing. Each leg is assigned with a transport mode and probably with a parameter trav\_time to describe the expected travel time for the leg. Leg will be directly executed after the last activity has ended. 

Parameter route is also required for execute simulation, but it can be different according to the modes. For example, route in car legs will list the links for agents to move along it in the given traveling direction, while route in transit legs will store the stop locations and expected transit services. Automatically calculation on initial routes for initial plans will be conducted by MATSim without containing them[7]. When the agent is behaving in the simulation, plans will be selected to conducted while only one plan can be executed at a time. Changing of plans occurs when replanning stage is going on. 

\textbf{Population-level data} is the features of the communities, such as the employment rate, income level, age distribution of residents, regional function (business, entertainment, university area, uptown or others) , which will influence the transits' tendency to some extent. This kind of data is usually obtained from the official census. Lazarus, Jessica et al., (2020) [11] added the population-level data - tract-level population, job count, employment rate, age, income, and gender distribution, which come from the U.S. Census (ACS, 2016; LEHD, 2013)- to the basic trip-level data as one of the input files. 

\textbf{Travel demand files} are needed for all the micro-mobility simulation models. Generally, census data, observational experiments, and stated-preference surveys are the common demand data resources. The datasets include the basic trip information such as the trip duration, start (or end) times and location (usually the coordinate of latitude and longitude, that is the origin and destination of the trip), as well as users' features in some studies. While some researchers found other resources of data for demand generation, like transport card transactions. In Lovric et al. (2013) generated the demand by smart-card transactions (individual check-in and check-out activities with occurrence time and location) data collected in a Dutch urban area, and inputted them into the model.  

\textbf{3D Modeling of micro-mobility vehicles} \textbf{\space }Deepak Talwar[10] modified the LGSVL Automotive Simulator to include micro-mobility vehicles by adding the 3D modeling of micro-mobility vehicles and extra components into the simulation environment. According to the popularity in urban society and appearance as well as motion models, five different kinds of micro-mobility vehicles were chosen to be simulated. 

\textbf{*summarize}

In short, all the variables, data and files which are put into the model for formulating the basic rules and framework can be collectively referred to as inputs. Basically inputs for model configuration, network designing, setting up the impact environment, demand generating and behavioral rules are required, while detailed required inputs will vary according to the applied methodologies.

\subsection{Variables}Traffic flow of micromobility, like any other traffic flow in the real world, is continuously changing and evolving. Actually, the evolution of road traffic is even more complex thanks to the interaction of various means of transportation. It's impossible to restore the whole traffic system completely. Because there're endless variables and the things will be different if any single variable changes. Simulating large road networks and enormous agents will be a nightmare when running time is taken into account. So finite representative variables must be chosen to reflect the real traffic as well as minimize the uncertainty and inaccuracy caused by ignoring the inconspicuous features of road traffic.

To find out the similarity of the variables, we first categorize them into the following six groups. Each of them contains some examples. On the other hand, as explained in taxonomy, parameters contained in the whole micro-mobility simulation can be generally divided into five categories according to the role played in the simulation.

\subsubsection{Related objectives} Then, we focus on different factor types and typical examples correspondingly to answer the following 4 questions:

Why are these factors chosen?

How are these factors related to their simulations?

How can these factors reflect the evolution of the model?

What principal should we follow to adopt these factors into our model?

\paragraph{Variables related to a person or a passenger}Subjective needs are one of the key elements to influence and even alter one's decision, thus leading to a series of changes to the transportation system. Therefore, it's necessary to visualize and quantify people's needs and put them into simulations as factors.

Wang I~L, et al. [1] simulate the service requests by setting the lower bounds of bike inventory as and the upper bounds as in order to find proper bike redistribution strategy. If the inventory level isn't within the bounds (i.e. ), a service request will be assigned. Only one service request is considered at a time. If there's a loading request (i.e. ), then bikes should be unloaded from vehicle to the site. Similarly, bikes should be loaded from the site to vehicle. and represent a safety bike buffer or empty rack buffer respectively. 

If more detailed information is considered, factors reflecting the trip and the access and excess condition should be included. For instance, travel and connect distance  and total bicycle access and egress time are used to represent the travel demand by Lu M, et al. [5] and Romero J P, et al. [7] respectively, because (i) they are most likely to influence passengers' mode choice; (ii) with the distance data, the interaction between travelers can be simulated.

Some other factors should be included for specific researches. Bicycle access time worth is considered by Romero J P, et al. [7] in order to meet their objective to take passengers' personal time value into account and then to achieve total efficiency and minimal social cost. Elhenawy M, et al. [8] regard the Home Index (HI) - the percentage of agents that finish their trips at home location {\textendash} as one of the two criterions to evaluate the behavior of their agent-based model when the central agent rules are changing, so they introduce the home location  as a factor to finally obtain the HI.

In conclusion, the PR factor in an agent-based simulation targeting MM should show individual's travel demands. The demands can be general or specific according to different objectives. Factors about the trip itself and its access and excess condition are widely used.

\paragraph{Variables related to vehicles}In agent-based simulations, vehicles and people are equally important. In most situations, they are both chosen as agents. On the other hands, vehicles have different features compared to people and MM is also unique to traditional vehicles like cars. In this part, we discuss how MM act as the role {\textendash} vehicle {\textendash} in transportation system and how to reflect them through factors.

Firstly, traffic flow data. Elhenawy M, et al. [8] record the current location of each vehicle every clock tick to check if its current location coincide with the origin of the trip to find if the vehicles (e-scooters in the paper) can fulfill the demand request. To reflect the evolution of the simulation, the demand generator gives all e-scooters new demand requests including information related to trips like location.

Secondly, factors about the vehicles itself. E-bike battery range [2][8] reflects the endurance mileage of the vehicle and it's crucial when allocating orders. When the e-bike is in used, the battery level will go down and when it's in a station, it can be recharged. The inclusion of battery level and other similar factors that shows the feature of vehicles can affect people's choice to some extent and thus make the simulation more accurate. For instance, users may favor bikes with higher battery level. Bikes with battery levels under a certain percentage will be showed unavailable.

Thirdly, factors that shows how the vehicles interact with other elements, such as the inventory level, the current request [1], the cost of travel [5], travel time [7], the number of available bikes [9]. Wang I~L, et al. [1] check if the inventory level is within the bounds to decide if they can accept the service request or not. Additionally, they also record the ID of the service request that the vehicle is serving currently in their simulation to better manage them. Bala\'{c} M, choose adoption rate as a factor corresponding to their long-term simulation. Changing of the adoption rate alter some outputs like access and egress time and rejection rate. Lu M, et al. [5] set the cost of travel as a factor since most MM companies use tiered price, so the travel bill vary from person to person and trip to trip. Whether the price is competitive enough is crucial to the survival of MM. Romero J P, et al. [7] include travel time to solve the planning problem to achieve the smallest social cost. Caggiani L, et al. use the number of available bikes reflect the situation of each station.

Fourthly, abstract factors that are used to adjust the model to make it more close to the reality. In MATSim's official guide [4], flow capacity factor and storage capacity factor are included. They should be set corresponding to different scenarios.

In conclusion, VH factors should be carefully selected and set for a satisfying simulation. Traffic flow data and factors about the vehicles itself are the basis since they reflect the role of MM in transportation system. Factors showing how the vehicles interact with other elements are also important when considering the evolution of the model. Abstract factors may sometimes be introduced to adjustment.

\paragraph{Variables related to roads}Road condition matters a lot in travel mode choice, efficiency and comfortableness, etc. Chen L, et al. include traffic events as a quantitative index about road condition. They record each type of traffic event on its nearby stations in the next hour after the traffic event occurs. These events may affect people's choice and the vehicles' running route.

Though road condition seems as an important factor, not many scholars take it into account. Some neglect it as they focus on macroscopic simulation and road condition is too micro for them. For those who do microscopic and mesoscopic simulation, perhaps they finally give up considering road condition because it takes more simulation time and compacity or they haven't found a good way to reflect road condition. If proper factors can be found or more powerful simulation engine is introduced, it's still necessary to take RD factors into account.

\paragraph{Variables related to context or surroundings}Context and surroundings can easily affect people's decision on transportation. One may not choose micromobiliy when it's raining or it's too hot. Chen L, et al. [6] select four factors related to context or surroundings {\textendash} date and time, weather condition, air temperature and social event. For the factor date and time, they first spilt two day types {\textendash} weekdays and weekends. Then, weekdays are divided into 4 parts - morning rush hours, day hours, evening rush hours and night hours; weekends are divided into day hours and night hours. Thanks to date and time factor, various bike usage pattern according to the time of the day can be observed and studied. As to weather condition, they define the following five groups: clear, cloudy, rain, snow, and haze. When it comes to air temperature, they split into below zero, cold, comfortable, and warm. 0\ensuremath{^\circ}C, 10\ensuremath{^\circ}C and 22\ensuremath{^\circ}C are set as boundaries.

\paragraph{Variables related to the condition of the origin and destination point}One may choose an origin point (O point) when there's enough bikes and the road around is bicycle-friendly; similarly, one may finish their trip (D point) in a station with an empty space or a place near a subway station. Therefore, the condition of the OD point is a factor that can't be ignored. Some factors are related to both O~point and D point (CO \& CD), while others only affect one point (CO or CD).

Most CO/CD factors are CO \& CD factors. Wang I~L, et al. [1] introduce the full rate of a rental site and categorize the sites into different levels according to it. However, the full rate of a rental site sometimes act differently in O~points and D points. =50\% or 100\% is almost the same for an O~point, but means it can't be a D point. In turn, =0\% or 50\% is almost the same for an D point, but means it can't be an O~point. So, should be distinguished according to different points though it's the same factor. Soriguera F, et al. [2] also care about whether the rental site is full, but they use it in a larger scale. They only care about full stations (marked 1) and empty stations (marked 0). Then the percentage of full and empty stations at a given instant is included in their simulation to get the peak value and low value of vehicle using. Bala\'{c} M, et al. [3] use a figure to show the location of all the pickup and dropoff stations. Lu M, et al. [5] include cost of connect mode to get a more accurate travel cost. It varies as time goes by due to the changing external environment. Romero J P, et al. [7] include number of docking stations and per-station cost  to calculate the minimum social cost. Elhenawy M, et al. [8] add the availability  of e-bike to the four state variables. The e-bike will be shown unavailable if someone else is using it or it has been rented. 

Some factors are only related to D points. In Wang I~L's simulation, [1] a better bike-return site, different from the preferred one, is specified by a commuter. If the bike-return site is expected to be full, then return the passenger will be suggested to return his bike at a nearest adjacent sites. Caggiani L, et al. [9] turn to the number of docking slots (free or full) as variables in their bike sharing system simulator to mark the whether the D point is available.

\subsubsection{The role in the model}

\paragraph{Configuration}Configuration parameters work for building the data structure, model framework as well as simulate the environmental factors of the system. As described in the Inputs part, configuration parameters will be defined and explained in the inputs files. Further subdivision yields two more categories: parameters for framework and environment.

Parameters for framework configuration build the structure of the system or model, ensuring simulation to behave with a settings list. For MATSim, they are usually pairs of a parameter name and a parameter value, determining the data structure and model infrastructure. And the settings of these parameters depends on the features of data sets used in studies. Some of them are selected from a variety of patterns, while some are assigned specific values. 

Parameters for environment endow the model with more details to promote the model has more realistic environment, and finally to gain more reliable results and solutions. Several aspects including infrastructure, social development, geographical elements and others could be taken into account. For infrastructure, number and locations of bicycle infrastructure(stations, lanes, etc)and other public infrastructure(parks, recreational area, commercial venues, industrial region, metro stations, bus stops, etc)are determined according the chosen cases. For social development, resident income, density and distribution of population, community and others can be taken into account. Temperature, elevation, slope, time period(morning peak period and evening peak period are commonly chosen), weekdays or weekends, weather, seasons and others are commonly considered geographical parameters. The specific settings depend on the cases.

\paragraph{Service level}Some parameters are considered in many studies to measure the efficiency and availability of the model, that is, customer service level. Coverage level, bicycle availability(i.e. existing available bikes in the station), bicycle stations availability(i.e. existing available bicycle docking for users to lock), maximum walking time or distance, traveling cost are mainly used measuring parameters to conduct behavior rules, estimate the simulation results and even optimize the facility distributions in cases. 

Likelihood of finding a bike/parking works when agents verify their next trip-activity by searching for available bikes at the origin station or docking racks in the destination station within the maximal radius. Cornelia Hebenstreita et al.[1] proposed a utility function to gain the probability, which includes three vital parameters: the intensity and direction of the influence quantity at alternative mode, the observed influence quantity of alternative mode for all agents, and the unobserved error term.

A similar parameter defined by Lin J-R et al.,[16] was named ``availability'' to measure the ratio of number of public bicycles available to station capacity. Therefore, the value of availability ranged from 0 to 1(from empty to full), and the stations were divided into five categories, among which the category with the lowest availability and the highest availability will be taken into account for rebalancing activities. 

Max walking distance or Max walking time is taken as a vital parameter by Miller et al.(1992) [14]. The value of walking from one node to at least one zone will be calculated to be assured to less than the given max walking time, for guaranteeing the walking worthwhile to substantially exist. 

Maximal distance is used to judge the agent choice by calculating the distances between origin and the nearest station and between destination and the nearest station for determining the start and end station. Cornelia Hebenstreita et al.[1] generated a preprocessed bike-sharing plan, and therefore the sequence transportation mode (walking, riding and taking a bus) will be decided according to the given distance limitation (maximal distance, access distance, and egress distance) and other parameters.

\paragraph{Network and trips}As described in the taxonomy above, these parameters act in setting up the transport network, trip generating and distribution. Multiple ways have been applied to address the network design problem by using different variables and assumptions. Commonly, the set of intersections and roads are denoted as vital elements of network model. Related constraints are denoted by parameters which are explicated in Part 2.3.2.4 (Constraint). For example, Duthie J et al.[17] formulated models which connect origin-destination(OD) pairs with bicycle paths while taking the bicycling service level and total cost into consideration at the same time. Another solution is Bi-level mathematical programming model, and at the lower level of which benefits to cars and bicycles as well as parameters related to travelers' behaviors or preferences will be considered for better determination. Romero, J. P. et al.[15] used a genetic algorithm to search for the distribution of a given number of docking stations, in order to maximize the number of users. Therefore, parameters contained in the iterative modal split and network assignment model, as well as other additional parameters, such as the network-related variables(bicycle lanes, number and locations of bicycle docking stations).

 Four-step transportation forecasting model(or urban transportation planning model, that is, UTP model) is commonly used to estimate the number of vehicles or people that will use a specific transportation facility. For example, the number of the vehicles on a planned road, the number of people visiting the specific bike stations to find an available bike, or the number of people reaching the destination station for parking. Firstly implemented in the 1950s at the Detroit Metropolitan Area Traffic Study and Chicago Area Transportation Study (CATS), the four steps are: trip generation, trip distribution, mode choice, and route assignment. 

In trip generation step, trip purpose, land uses, household demographics, and other socioeconomic parameters are set for determining the frequency and locations of origins or destinations of trips. 

In trip distribution step, origins and destinations generated in the last step are matched in this step. Generally there are two basic methods: growth factor methods and synthetic methods[13]. Constant factor method, average factor method, Fratar method and Furness method are typical growth factor methods, which are all older methods to conduct trip distribution work. Commonly used synthetic methods are gravity model (that is, an entropy maximizing model) and destination choice models. Gravity model applies to mono-centric urban regions with little attention on accessibility to transit, therefore it's not appropriate for many urban cases which has many dominant attraction regions and various types of transits such as intown-to-surburban, intown-to-intown and surburban-to-surburban trip flows. However, destination choice models overcome these limitations by applying proper specifications of utility to assure logically responding to changes in levels of service and demand. The following table describes the parameters consideration in the models:

\paragraph{Constraint}Time worth is the traveling cost per unit of time. Most of the simulation model try to minimize the total traveling cost, while each travel cost could be obtained by multiplying the time worth factor and the time. Miller et al.[14] set three time worth parameters: bicycle travel time worth, bicycle access time worth, and car travel time worth. While Parameters related to travelers' behaviors or preferences will be considered in the lower level for better determination.

Cost  is a vital factor which is considered in many ways for micro-mobility simulation. It's commonly taken account in the network design model, route choice model, traveler behavioral rules, utility functions, etc. Cornelia Hebenstreita et al.[1] took the marginal cost of traveling into consideration, which consists of time dependent costs, route determined costs and monetary costs. Denoted by Miller et al.[14], social cost consists of three parts: bicycle travel costs, total station cost and car travel cost. Given that the inputted parameter time worth(e.g. the traveling cost per unit of time) is determined, the values of the three cost parameters could be obtained and changed according to the results from the simulation. For instance, bicycle travel cost is the sum of the product of total bicycle travel time and bicycle travel time worth as well as the product of total bicycle access and egress time and bicycle access time worth. At the upper level of Bi-level programming model, cost(including social, economic, environment, etc) is considered to optimize the decision. 

Charging/discharging rules  should be considered in e-bikes simulation framework. In most of the literature, a simplified linear loss is generally used as the discharging energy flow curve, because the various influential aspects (battery capacity, riding speed, weight of rider, weather, gradients and others) in reality could complicate the calculation. For the charging rule, some of the studies will still use a linear curve while others proposed their functions to depict the changing of the energy. 

Minimal distance can be the given distance between one node to the nearest docking station in the study of Miller et al.(1992) [14], who didn't consider the cases that have docking stations closer than the minimal distance, in order to further make a reduction on the search space. In the reference [14], another minimal distance was also proposed to restrict that location of docking stations that any two of them cannot be closer than the minimal distance. 

Population size  were set differently in the outer Genetic Algorithm with the same number of docking stations to find the maximal number of public bicycle users[14]. 

Agent memory size is set to constrain the number of plans, and only the plans with the highest probability will be kept if the limit is exceeded to generate improving correlations. This parameter will act in the replanning process. Dubernet et al.(208;15)[4] 

User group provides the attitudes for the aspects in the riding, such as the routing choice(safety, smooth, gradient, comfort, congestion, environment, other). In MATSim, a user group and a bicycle type will be assigned to an agent. Cornelia Hebenstreita et al.[1] evaluate for all attribute groups from 0-10 to develop the parameter named bike-friendness, which can be used to generate the parameter called personal weighting for choosing the best route.

\paragraph{Utility}Utility functions are used to evaluate the performance of operation, while parameters and factors concerned in current studies are different. Standard MATSim utility function[3](also named MATSIM scoring function) can be used to determine whether the equilibrium is reached. Agents will gain utility when they are performing activities and dis-utility when they are traveling. It was firstly developed by Charypar and Nagel(2005) on the basis of the Vickrey model for road congestion, contains the definition of the activity utility and travel utility. The basic function considers activity duration, waiting time, as well as penalties of late arrival, not staying long enough and very short trip. All the parameters included in the basic function are explained in Table x. 

Table x 

|Table 3:|||parameter ||explanation ||parameter ||explanation || ||the utility of an activity || ||the performed activity duration || ||travel dis-utility for a leg || ||related to the marginal utility of activity || ||||||the duration when utility starts to be positive || ||the waiting time spent || ||the direct marginal utility of time spent waiting || ||the late arrival penalty || ||the waiting time || ||the penalty for not staying long enough || ||the activity starting time || ||the penalty for a very short activity || ||the latest possible penalty-free activity starting time || ||the activity ending time || ||the earliest possible activity end time || ||related to the penalty for not staying long enough || ||the shortest possible activity duration || ||a mode-specific constant || ||the direct maginal utility of time spent traveling by mode || ||the travel time between two activity locations || ||the marginal utility of money ||\ensuremath{\Delta } ||the change in monetary budget || ||the marginal utility of distance || ||the mode-specific monetary distance rate || ||the distance traveled between activity || ||public transport transfer penalties || ||a 0/1 variable signaling whether a transfer occurred ||

Although the original purpose of the scoring function was to choose departure time, further choice dimensions have been verified to successfully realize. In order to conduct more reasonable simulation, many properties of riding were taken into account to extend the standard function. Dominik Ziemkea et al.[2], who firstly proposed the concept of person dependent perception of trip duration[1], added an infrastructure and a comfort component to revise the travel utility, considering the cycling infrastructure, pavement conditions and riding smoothness. Felt travel time (as) was then introduced by Cornelia Hebenstreita et al.[1] acting as one of the scoring for riding to deal with different extent of preference among factors (route, comfort and so on). It has two sub-parameters, that are the felt travel times of each edge and the marginal costs of traveling. In addition, some studies conducted the integration of social networks in multi-agent simulation of MATSim framework. Hackney et al.[5] proposed their formulation of socially-networked utility specifications, that joint utility function allows the attributes, utility as well as attitudes of at least one agent to act as decision variables in the utility of another agents. Therefore, they attached attention to social relationships, and four social interaction functions were explicated. Instead of only focusing on intrahousehold ties, Dubernet et al.[4] designed a model containing explicit coordination of individuals to represent joint decisions, which can deal with complicated network topology. They added two parameters to the basic function, a random utility , for each activity to express the preference variations; and the other, the utility of time passed with sicial contact. 

*Summary

On the whole, five kinds of parameters involved in micro-mobility simulation can be found. Configuration parameters contained in the inputs will determine the data structure and framework of the model; service level-related parameters service for representing the travel indicators of the system as a judgment of whether the system runs well; parameters in network and route designing process construct the structure of the transportation network and formulate the route selection behavior rules; parameters for constraints will set the limitation of some of the values or assigning quantifiable values to an otherwise unquantifiable parameter; parameters works in utility functions will contribute to evaluating utility and performance of the system. What calls for special attention is that parameters are the basic elements of the simulation, and therefore are partly involved by inputs. Some of the other parameters do not need to input, for the implementation of the simulation will assign them values.

\subsection{Calibration methods {\textemdash} What were the methods that were applied for calibration? How were they different from each other? How were their usability related to the inputs or parameter settings? Follow the above organization}Compared  with traditional macroscopic models, microscopic models are more  commonly used to study traffic  flows in smaller areas where calibration is  required due to the increased  number of model parameters (Fellendorf \& Vortisch, P, 2001). Numbers of parameters were used to validate against field data. To find out the similarity of the validation metrics, we divided them into two categories: traffic flow characteristics and passenger behavior.

\begin{table*}[!htbp]
\caption{{} }
\label{tw-f2eab6a961f4}
\def\arraystretch{1}
\ignorespaces 
\centering 
\begin{tabulary}{\linewidth}{LLL}
\hline 
\multicolumn{3}{p{\dimexpr(\mcWidth{1}+\mcWidth{2}+\mcWidth{3})}}{Table x: The classification of validation metrics}\\
Metrics type &
  Abbreviation &
  Example(s)\\
Passenger/driver behavior &
   &
  standstill longitudinal,  \mbox{}\protect\newline distance between the stopped vehicles,  \mbox{}\protect\newline headway time in seconds, \mbox{}\protect\newline Contact duration(time two visitors remain in contact), \mbox{}\protect\newline number of neighbors, \mbox{}\protect\newline Inter-contact time (ICT) is the time elapsed between starts of two successive contacts of the same pair of  \mbox{}\protect\newline devices/visitors. \mbox{}\protect\newline Inter-contact time (ICT) is the time elapsed between starts of two successive contacts of the same pair of  \mbox{}\protect\newline devices/visitors. \mbox{}\protect\newline Inter-contact time(the time elapsed between starts of two successive contacts of the same pair of visitors), \mbox{}\protect\newline The percentage of new contacts \\
traffic flow &
  VH &
   The road network flow capacity, \mbox{}\protect\newline Transit vehicle capacity/volume, \mbox{}\protect\newline the unsaturated speed, \mbox{}\protect\newline Trip mode share, \mbox{}\protect\newline Tour mode share, \mbox{}\protect\newline Trip length frequency, \mbox{}\protect\newline Volume of trips, \mbox{}\protect\newline Trip distribution, \mbox{}\protect\newline Queue length, \mbox{}\protect\newline Total delay time,  \mbox{}\protect\newline Total travel time,  \mbox{}\protect\newline Cost of total delay \\
\hline 
\end{tabulary}\par 
\end{table*}

\begin{itemize}
  \item \relax Calibration metrics related to passenger (PR)
  \item \relax Calibration metrics related to vehicles (VH)
\end{itemize}
  
\begin{table*}[!htbp]
\caption{{} }
\label{tw-c3c23eda2b4b}
\def\arraystretch{1}
\ignorespaces 
\centering 
\begin{tabulary}{\linewidth}{LLLLL}
\hline 
 &
  \unskip~\cite{1219518:23250172} &
  \unskip~\cite{1219518:23250173} &
  \unskip~\cite{1219518:23250170} &
  \\
frequency of no battery &
  \ensuremath{\surd }  &
  \ensuremath{\surd }  &
  \ensuremath{\surd }  &
  \\
frequency of no bicycle &
  \ensuremath{\surd }  &
  \ensuremath{\surd }  &
  \ensuremath{\surd }  &
  \\
service rate &
  \ensuremath{\surd }  &
   &
   &
  \\
margin cost &
   &
  \ensuremath{\surd }  &
   &
  \\
felt travel time &
   &
  \ensuremath{\surd }  &
   &
  \\
waiting list handling &
   &
  \ensuremath{\surd }  &
   &
  \\
station selection &
   &
  \ensuremath{\surd }  &
   &
  \\
derivation of bs types &
   &
  \ensuremath{\surd }  &
   &
  \\
access time(distance) &
   &
   &
  \ensuremath{\surd }  &
  \\
frequency of full stations(can not lock) &
   &
   &
  \ensuremath{\surd }  &
  \\
leave for empty stations &
   &
   &
  \ensuremath{\surd }  &
  \\
\hline 
\end{tabulary}\par 
\end{table*}
Agent-based (AB) transport models simulate the travel demand at the scale of \textbf{individuals}. In general, this level of detail requires \textbf{higher computational capacity} and \textbf{longer runtimes}, compared to aggregated models that simulate traffic flows. To run large-scale AB scenarios in reasonable times, the use of \textbf{small subsamples} of the population (down sampling) is common practice\unskip~\cite{1219518:23250168}. However, down sampling will definitely lead to uncertainty\unskip~\cite{1219518:23250167}. How to keep balance between a proper scale and certainty is critical to multi-agent simulation.\textbf{\space }

\textbf{5\% of agents and 40-50 iterations} are suggested in a scenario where results are analyzed only at \textbf{a highly aggregate level\unskip~\cite{1219518:23250168}\unskip~\cite{1219518:23250166}. }Admittedly, the requirements of the policy scenarios may still affect the most suitable scale factor and network density for each application. For detailed analysis of small areas, the random effects across simulation runs should be also taken into account\unskip~\cite{1219518:23250167}. The effects of scaling on public transport trips require further attention, since vehicle capacities cannot be scaled down in the same proportion.

So, if the computing power is allowed, we can set the agent rate at \textbf{30\%}, in order that the occupancy of the public transit aren't adversely impacted\unskip~\cite{1219518:23250165}. The suggested rate can be summarized in the following table:

\begin{table*}[!htbp]
\caption{{} }
\label{tw-3ee11661bd83}
\def\arraystretch{1}
\ignorespaces 
\centering 
\begin{tabulary}{\linewidth}{LLL}
\hline 
Suggested agent rate &
  Population sample  &
  Public transit included\\
5\% &
  Small (2.9 million total trip per day) &
  No\\
30\% &
  Large (3.5 million private cars \& 1.7 million public vehicles) &
  Yes\\
\hline 
\end{tabulary}\par 
\end{table*}
Additionally, the adverse impact of population sampling is determined mostly by the patterns of the agents' behavior rather than by the traffic model. So, we'd better focus on \textbf{the patterns of the agents' behavior}.

\begin{table*}[!htbp]
\caption{{} }
\label{tw-7ba8d95c4db1}
\def\arraystretch{1}
\ignorespaces 
\centering 
\begin{tabulary}{\linewidth}{p{\dimexpr.09040000000000001\linewidth-2\tabcolsep}p{\dimexpr.3096\linewidth-2\tabcolsep}p{\dimexpr.20\linewidth-2\tabcolsep}p{\dimexpr.20\linewidth-2\tabcolsep}p{\dimexpr.20\linewidth-2\tabcolsep}}
\hline 
 &
  \multicolumn{2}{p{\dimexpr(.5096\linewidth-2\tabcolsep)}}{sensitive analysis} &
  \multicolumn{2}{p{\dimexpr(.40\linewidth-2\tabcolsep)}}{linear regression analysis}\\
\multicolumn{1}{p{\dimexpr(.09040000000000001\linewidth-2\tabcolsep)}}{\multirow{2}{\linewidth}{\unskip~\cite{1219518:23250172}}} &
  \multicolumn{2}{p{\dimexpr(.5096\linewidth-2\tabcolsep)}}{activity duration(trip length) - demand - min bikes(batteries) needed} &
  \multicolumn{1}{p{\dimexpr(.20\linewidth-2\tabcolsep)}}{\multirow{2}{\linewidth}{Activity duration has a strong effect on the number of e-bikes and batteries needed to serve the demand.  \mbox{}\protect\newline R2=0.83min e-bikes required \mbox{}\protect\newline R2=0.84min batteries required}} &
  \multicolumn{1}{p{\dimexpr(.20\linewidth-2\tabcolsep)}}{\multirow{2}{\linewidth}{Battery requirements are about twice as sensitive to trip length as e-bike fleet requirements.}}\\
 &
  a strong positive relationship between the minimum number of e-bikes/batteries required and the trip rate &
  A correlation between the minimum number of e-bikes/batteries and the trip length is less evident. &
   &
  \\
 &
   &
   &
   &
  \\
 &
   &
   &
   &
  \\
\hline 
\end{tabulary}\par 
\end{table*}

\subsection{Comparison to other models}Compared to Equation-Based Models(EBMs), Agent-Based Models(ABMs) are more detailed, with higher accuracy\unskip~\cite{1219518:23250164}, and can complement EBMs. ABMs allows further exploration of macroscopic implications based on lab experiments, generating new hypotheses, and determining the sensitivity of the results. Akka (which we decide to apply) is one of important actor(agent) models. It is feasible, reliable and scalable, and Akka dispatcher can be used to iterate over multi-scale application modules\unskip~\cite{1219518:23250163}.

\begin{table*}[!htbp]
\caption{{} }
\label{tw-6a56c7d34323}
\def\arraystretch{1}
\ignorespaces 
\centering 
\begin{tabulary}{\linewidth}{LLLLLL}
\hline 
Category &
  Performance &
  Speed &
  Maximum number of nodes cluster &
  Model size &
  Reference\\
\multicolumn{1}{p{\dimexpr(\mcWidth{1})}}{\multirow{3}{\linewidth}{Control and monitoring of agent-based networks}} &
  The largest gain in performance is about 31\% &
  250 frame per second(FPS) &
  1000 &
  2400 cells &
  \unskip~\cite{1219518:23250162}\\
 &
  average CPU utilization is 10\% &
  15 to 30 seconds to add 20 nodes &
  2400 &
   &
  \\
 &
  High efficiency in CPU utilization &
  an average of 11 minutes on 144 nodes (3456 cores). &
  2232(53,568 cores) &
  randomly populated grid of size 5.20 \ensuremath{\times} 107 &
  \unskip~\cite{1219518:23250161}\\
Virtualization for control and monitoring &
  3\% improvement in deduplication rate &
  89.76\% overall reduction in migration time &
   &
  83\% reduction in the overall storage of images &
  \unskip~\cite{1219518:23250160}\\
Control and monitoring other networks via akka &
  In the 10-iteration case, this slows the job down by 50s (21\%) on average. &
  the analytics is done over 10 minutes, sensors collect readings every 10 seconds &
  8000 &
   &
  \unskip~\cite{1219518:23250159}\\
\hline 
\end{tabulary}\par 
\end{table*}

\section{Impacts/Application Fields}

\subsection{Overview of the applications for micromobility simulation (Energy, system design, system optimization, demand prediction, etc..) }The rising of micro-mobility services over the past decade promotes an emerging trend in studies and developments regarding design and management methods. The idea of improving the efficiency of a system is to better allocate resources following the demand. Hence, existing studies concerning the operation of a micro-mobility service mainly focus on either the supply- or the demand-side of the reality. Based on how it manages to cooperate real-world data with its research purpose, these studies could be categorized mainly into analytic and simulation-related topics. Analytic studies utilize data for static pattern analysis through methods such as modeling, Spatio-temporal analysis, etc. These help to discover potential long-term or periodic interactions between demand and supply concerning internal (e.g. user socio-demographics), external (e.g. weather, artificial environment), and trip-related (e.g. origin, destination, distance, time, and date) factors\unskip~\cite{1219518:23250149}. However, short-term prediction and analysis are more important for real-time management. Studies on simulation of micro-mobility services focus on capturing dynamics of system demand and supply by modeling the behavior of each element within the system, which is more relevant to the topic of this paper. 

Simulation has been developed and utilized as a tool in the design and management of transportation infrastructure for decades. Based on the scales, they could be either microscopic, mesoscopic, or macroscopic simulations. Well-known tools of which include SUMO \citep{lopez2018microscopic} , MATSim\unskip~\cite{1219518:23250147} , and PTV Visum. However, as an emerging mobility service, simulations of the micro-mobility system are now getting attention from the research community across the globe. The simulation normally takes information about O-D or trips of users as inputs, presets a series of parameters that help depict user behavior (e.g., mode choices), and generates outputs for analysis.

Some of the studies in this area built models from a macroscopic perspective, where models consider users from a large scale area with normally static O-D demands and route choices. Caggiani et al.\unskip~\cite{1219518:23250146} first proposed a simulator for validating an algorithm of reposition. It takes preset pickup demand as inputs for each discrete time interval, derivesdestination choice based on relative O-D attractiveness and the nature of the trip, removesusers who fail to hire a bike, and sets those who fail to return the bikes as waiting. The model was later used and tested in a subsequent study\unskip~\cite{1219518:23250145}  on proposing a dynamic simulation-based fleet reposition algorithm. In 2014, Ji et al.\unskip~\cite{1219518:23250172}  evaluated the efficiency ofthe shared e-bike systems through a proposed Monte Carlo simulation model. The model consists of a trip generation module and processing events, which handle demand generation and the arrival, return, and recharging of e-bikes, respectively. But the paper only considered round trips, and the trip demands and round trip length were randomly derived from distributions instead of real-world data. Similar approaches are seen in other studies\unskip~\cite{1219518:23250143,1219518:23250142,1219518:23250141}. Despite the rather low computing costs and sufficient precision, macroscopic models fail to depict the behavioral properties of the elements within a system. Many of the above-mentioned studies utilize local data to calibrate a distribution, which only applies to the condition of the data source. Besides, the static O-D demands and route choices can't describe situations where users switch to other transport modes during the trip.Mesoscopic simulation, which takes behaviors and dynamics of elements into account while keeping the simplicity by clustering different users, has become a widely adopted alternative in recent years. And agent-based models (ABM) and multi-agent simulation rise amongst the most popular simulation approaches. In 2018, Hebenstreit et al.\unskip~\cite{1219518:23250140} developed two modules for MATSim agent-based simulation, which enable a simulation of bicycle traffic and bike-sharing within a multimodal context. The paper wasone of the earliest studies that consider user choices and modal shifts. Soriguera et al.\unskip~\cite{1219518:23250139}  presented a comprehensive simulation model based on Matlab to support the planning and design of a bike-sharing system. Repositioning trucks and electric bikes alternatives were included in this model. The other two subsequent studies[13],[14] also explored the potentials of ABMs with different settings of agents and parameters using data from BiciMAD. In short, the above-mentioned works on micro-mobility simulation mainly focus on the balancing problem within micro-mobility systems. The mesoscopic models have higher complexity for computation, but could better capture dynamics of users regarding modal splits and route choices during the trips. Nevertheless, most of the existing studies research on a local area with small amounts of trip inputs. And they lack a systemic analysis for energy efficiency. Hence, the proposed simulation model in this paper highlights the energy efficiency of the micro-mobility system. Incorporated with Akka distributed computing framework, the proposed method enables fast processing of a mesoscopic model with concrete trip plans (e.g. home-based trip, work-based trip, non-work-based trips, etc..) and derives energy and emission outputs for further analysis

The Behavioral Energy Autonomous Mobility Model (BEAM) is a microscopic agent-based travel demand simulation framework based on the Multi-Agent Mobility Simulation Model (MATSim) \unskip~\cite{1219518:23250151} . Agents are measured by traveling time, expense or costs and choice preferences. Successor to MATSim, BEAM incorporates the AKKA library aiming to perform scalable analysis of massive amounts of data from urban transportation systems, in particular to improve the analysis of the benefits of plug-in electric vehicles (PEVs)\unskip~\cite{1219518:23250150} . BEAM outperforms MATsim in three ways. First, BEAM's mode choice is exogenously defined. BEAM allows for multi-modal choices, including walking, biking, driving alone, riding in a car, and a variation of walking-riding-driving. Secondly, BEAM's behavioral focus is intra-day planning, which means that BEAM can better reflect and simulate modern transportation. Finally, BEAM specifies a power threshold range constraint for electric vehicles.



%

\bibliographystyle{IEEEtran}

\bibliography{article}

\end{document}